\documentclass{emulateapj}
\submitted{Accepted by ApJL} 

\newcommand{\uv}{$u$-$v$}

\begin{document}
\newcommand{\kms}{\,km\,s$^{-1}$}
\newcommand{\mjb}{\,mJy\,beam$^{-1}$}

\title{First VLBI Detection of the Radio Remnant of Supernova 1987A:
Evidence for Small-scale Features}
\shorttitle{VLBI Detection of SN 1987A}

\author{C.-Y.~Ng\altaffilmark{1,2,a}, T.~M.~Potter\altaffilmark{3},
L.~Staveley-Smith\altaffilmark{3,b}, S.~Tingay\altaffilmark{4,b},
B.~M.~Gaensler\altaffilmark{2,c}, C.~Phillips\altaffilmark{5},
A.~K.~Tzioumis\altaffilmark{5} and G.~Zanardo\altaffilmark{3}}
\altaffiltext{1}{Department of Physics, McGill University, Montreal, QC H3A 2T8, Canada}
\altaffiltext{2}{Sydney Institute for Astronomy, School of Physics, The University of Sydney, NSW 2006, Australia}
\altaffiltext{3}{International Centre for Radio Astronomy Research (ICRAR) --
The University of Western Australia, Crawley, WA 6009, Australia}
\altaffiltext{4}{ICRAR -- Curtin University, Bentley, WA 6102, Australia}
\altaffiltext{5}{Australia Telescope National Facility, CSIRO, Marsfield, NSW 1710, Australia}
\altaffiltext{a}{Tomlinson Postdoctoral Fellow}
\altaffiltext{b}{Western Australian Premier's Fellow in Radio Astronomy}
\altaffiltext{c}{Australian Research Council Federation Fellow}

\email{ncy@hep.physics.mcgill.ca}

\begin{abstract}
We present a detailed analysis of the first very long baseline
interferometry (VLBI) detection of the radio remnant of
supernova 1987A. The VLBI data taken in 2007 and 2008 at 1.4
and 1.7\,GHz, respectively, provide images sensitive to angular
scales from 0\farcs1 to 0\farcs7, the highest resolution to date at radio
frequencies. The results reveal two extended lobes with an
overall morphology consistent with observations at lower
resolutions. We find evidence of small-scale features in the
radio shell, which possibly consist of compact clumps near the
inner surface of the shell. These features have angular extent
smaller than 0\farcs2 and contribute less than 13\% of the
total remnant flux density. No central source is detected in
the VLBI images. We place a 3$\sigma$ flux density limit of
0.3\,mJy on any pulsar or pulsar wind nebula at 1.7\,GHz.
\end{abstract}


\section{Introduction}
\object[SN 1987A]{Supernova (SN) 1987A} in the Large Magellanic Cloud
is the only nearby
core-collapse supernova observed in the age of modern astrophysical
instrumentation and has been intensively studied over the last
two decades. Optical observations reveal a triple-ring nebula
centered on the supernova, as part of the circumstellar material
(CSM) expelled by the progenitor star in its late evolution stage
\citep{ch91}. With a radius of 0\farcs86, the nebula's inner ring
marks the inner surface of the dense CSM \citep{plc+95}.
Images from the \emph{Hubble Space Telescope (HST)} show hot
spots emerging from the inner ring since 1995 \citep{lsb+00}.
There are currently about 30 optical hot spots encircling the
entire ring \citep{fmh+10}, which are believed to be clumps
of inward-protruding CSM shocked by the supernova blast wave
\citep{pmz+02}. In recent years, the interaction between the
blast wave and the inner ring has led to a rapid increase in the
radio and soft X-ray emission \citep{zsb+10,pbg+07}, providing a
good opportunity to study the shock interaction with the CSM.

At radio frequencies, the initial outburst was detected with
the Molonglo Observatory Synthesis Telescope \citep{tcb+87},
the radio flux then decayed rapidly and re-emerged in mid-1990,
marking the birth of a radio remnant \citep{tcm+90}. Since then,
the source has been monitored regularly with the Australia Telescope
Compact Array (ATCA) \citep[see][and references therein]{ngs+08,zsb+10}.
Super-resolved images at 9\,GHz reveal a two-lobe structure of
the radio shell \citep{sbr+93,gms+97}, which was later confirmed
with diffraction-limited images at higher resolutions at 18 and
36\,GHz \citep{mgs+05,psn+09}.

High spatial resolution is the key to study the supernova shock
evolution and its interaction with the CSM. The first very long
baseline interferometry (VLBI) observations of SN~1987A were made
5.2 days after the explosion. Although the radio emission was
completely resolved and hence no detection was made, the
observations posted a lower limit of $1.9\times 10^4$\kms\ on
the expansion velocity \citep{jkb+88}. A subsequent attempt
at VLBI in 2003 September also resulted in a non-detection
(J.\ Lovell \& R.W.\ Hunstead 2007, private communication).
The first VLBI detection of the radio remnant of SN~1987A was
made with the Australian Long Baseline Array
(LBA)\footnote{\url{http://www.atnf.csiro.au/vlbi/overview/}}
in 2007, and initial results have been reported by
\citet{tpa+09}. In this Letter, we present a detailed analysis
of the 2007 dataset, together with new observations in 2008,
and present evidence of small-scale features in the radio shell.

\section{Observations and Data Reduction}

\begin{figure}[th]
\epsscale{0.8}
\plotone{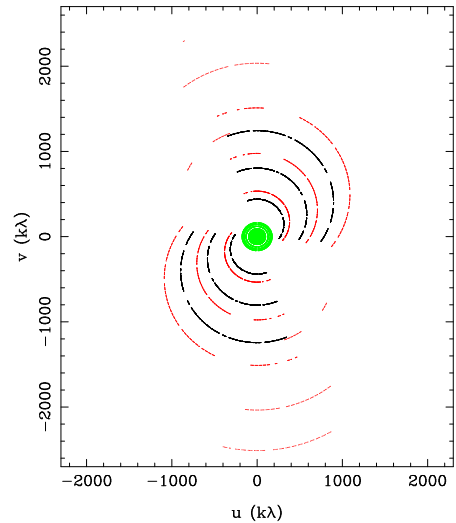}
\caption{\uv\ coverage of 2007 (black) and 2008 (red) VLBI 
observations at 1.4 and 1.7\,GHz, respectively, comparing to
that of 9\,GHz ATCA observations (green). \label{uv}}
\end{figure}

VLBI observations of SN~1987A were carried out on 2007 October 7
and 2008 November 26 at 1.4 and 1.7\,GHz, respectively, with
about 10 hours integration time each, using part of
the LBA including the ATCA ($4\times22$\,m antennas used as a
tied array in 2007, and 5 antennas in 2008), Parkes (64\,m) and
Mopra (22\,m) telescopes of the Australia Telescope National
Facility. The 2008 epoch has additional data taken from the NASA
DSS~43 antenna at Tidbinbilla (70\,m) for three hours to boost the
\uv\ coverage and sensitivity. The observation parameters are
listed in Table~\ref{table}, and the \uv\ coverage is plotted
in Figure~\ref{uv}. The 2007 and 2008 data are centered on
1.382 and 1.666\,GHz, respectively, with four dual-polarization
intermediate frequencies and a total bandwidth of 64\,MHz divided
into four bands (except from the Tidbinbilla antenna, where only the left
circular polarization is available). A nearby calibrator, PKS
0530$-$727, was observed every 10 minutes for phase referencing
purposes.

\begin{deluxetable}{lll}
\tablecaption{Observation and best-fit model parameters for 
2007 and 2008 VLBI observations of SN~1987A\label{table}}
\tablewidth{0pt}
\tablecolumns{3}
\tablehead{\colhead{Parameter} & \colhead{2007} & \colhead{2008}}
\startdata
\multicolumn{3}{c}{Observation Parameters}\\\hline
Date & 2007 October 7 & 2008 November 26\\
Day since explosion & 7531 & 7947\\
Center frequency (GHz) & 1.382 & 1.666\\
No. of stations & 3 & 4\\
Integration time (hr) & 9 & 10\\
Baseline (k$\lambda$) & 260$-$1250 & 300$-$2510\\
Baseline-hour & 27 & 36\\\hline
\multicolumn{3}{c}{Best-fit Truncated-shell Parameters\tablenotemark{a}}\\\hline
Radius (\arcsec) & $0.82\pm0.01$ & $0.84\pm0.01$\\
Half-opening Angle (\arcdeg) & $23\pm4$ & $18.2\pm1.2$\\
Thickness (\%) & $25\pm8$ &  $17.7\pm1.4$\\
$\chi^2_\nu/\nu$ & 0.81/25848 & 0.61/5122
\enddata
\tablenotetext{a}{obtained with $\chi^2$ (i.e.\ weighted)
fits; difference between the weighted and unweighted
fits give the reported uncertainties.}
\end{deluxetable}

The 2007 observations were made as part of an Express Production
Real-time e-VLBI Service (EXPReS) demonstration of a global
real-time interferometer. The digital data for four
dual-polarization 16\,MHz bands were transferred in real time
from each antenna in Australia via a dedicated 1\,Gbps lightpath
to the correlation facility on the opposite side of the Earth
(Joint Institute for VLBI in Europe [JIVE] in the Netherlands),
using standard gigabit Ethernet network adapters and switches
within the observatories. All four stokes parameters were
correlated at the European VLBI Network (EVN) MarkIV correlator
at JIVE. Technical details on the lightpaths and correlation
process have been reported by \citet{tpa+09}. The 2008 observations
were performed in regular VLBI mode where the voltage data was
recorded on disk, and subsequently correlated
at the Curtin University facility.

All post-correlation calibration and processing were carried
out with the AIPS package. Images were formed with natural
weighting and then deconvolved using the \emph{CLEAN} algorithm
\citep{cla80}. We performed self-calibration with a three-minute
timescale. Since the calibrator delivers high quality phase
corrections, this offers only a minimal improvement to the images.
Figure~\ref{img}a shows the final intensity maps
of the 2007 and 2008 observations, which have rms noise of
0.35\mjb\ and 0.1\mjb, and restoring beams of FWHM $0\farcs17 \times
0\farcs08$ and $0\farcs12 \times 0\farcs05$, respectively.

\section{Results and Discussion}
\subsection{VLBI Results and Comparison with Observations at
other Frequencies}

\begin{figure*}[ht]
\begin{center}
\epsscale{0.85}
\plotone{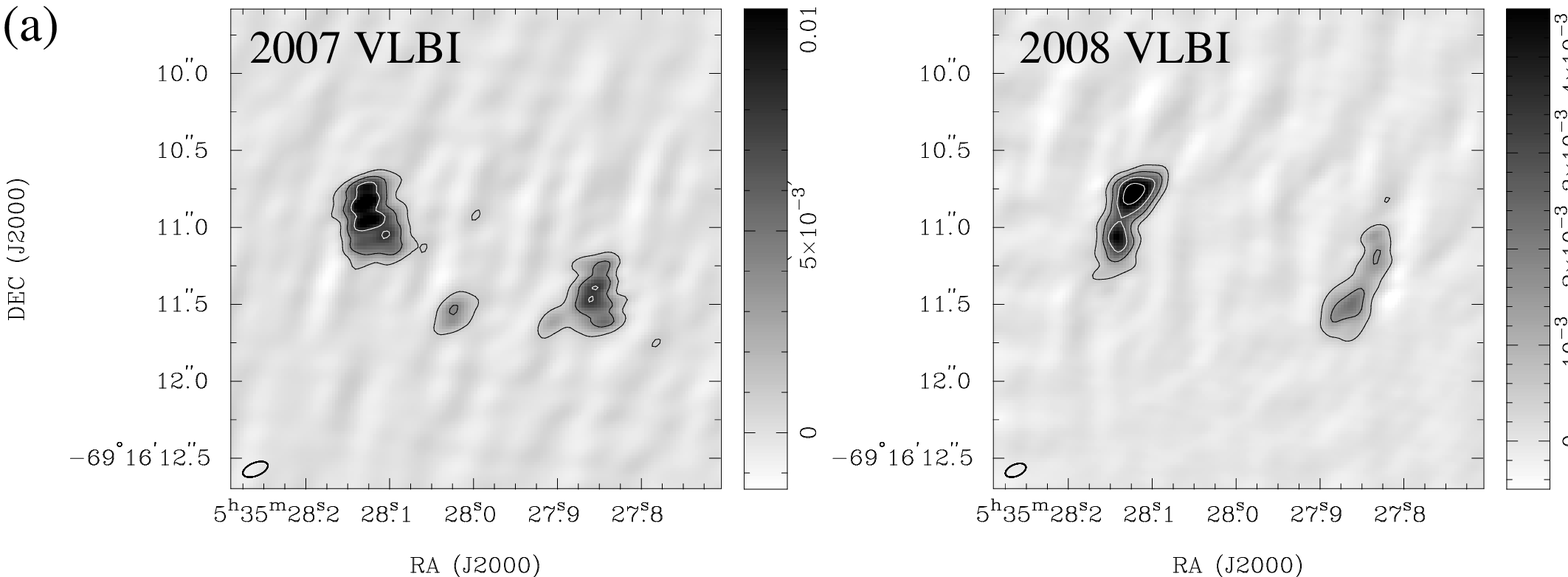}
\\[5mm]
\epsscale{0.7}
\plotone{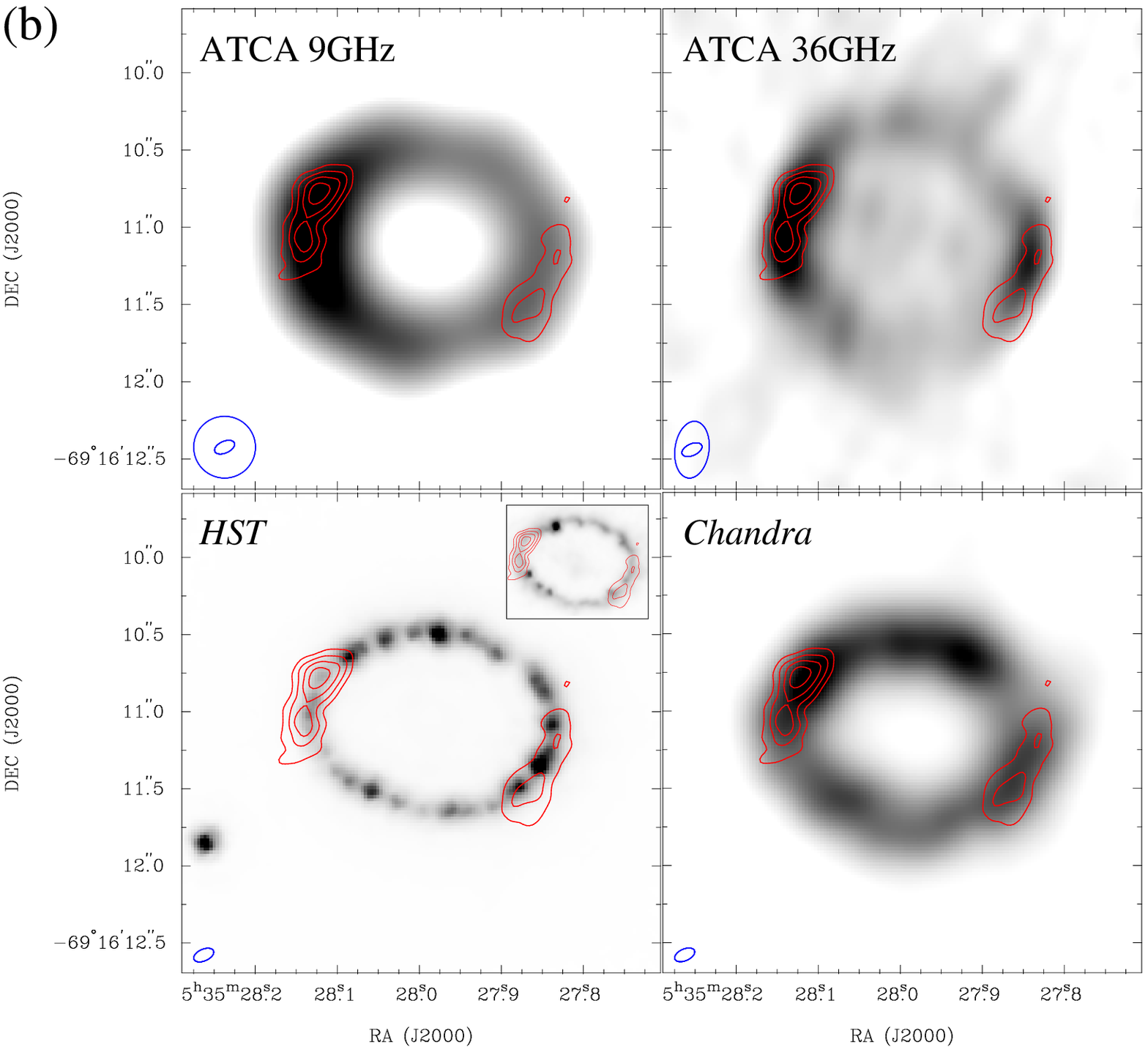}
\end{center}
\caption{(a) VLBI images of SN~1987A at 1.4 and 1.7\,GHz,
taken on 2007 October 7 and 2008 November 26, respectively.
The contours correspond to 1, 4, 8\mjb\ in 2007, and
0.5, 1.5, 3 and 5\mjb\ in 2008. The restoring beams are shown
in lower left, and the scale bars are in units of
Jy\,beam$^{-1}$.
(b) Comparison of the 2008 VLBI results to:
super-resolved 9\,GHz ATCA image taken on 2008 October
   11 \citep{ngs+08},
diffraction-limited 36\,GHz ATCA image taken on 2008
   October 7-12 \citep{psn+09},
\emph{HST} H$\alpha$ images taken on 2007 May 11 and
   2000 November 14 (inset), and
deconvolved \emph{Chandra} X-ray image in $0.3-10$\,keV band,
   taken on 2008 April 28-29 \citep{ngm+09}.
\label{img}}
\end{figure*}

The VLBI images of SN~1987A in Figure~\ref{img}a clearly
resolve the remnant structure and show a two-lobe morphology.
The lobes are separated by 1\farcs5 and are each extended with a
size $\sim0\farcs6\times0\farcs3$, a few times larger the restoring
beams. The eastern lobe has a higher flux density than the western
one by a factor of 2.5. There is some hint that the lobes
consist of distinct peaks. A comparison between the 2007 and 2008 images
show a similar overall morphology, but the lobes in the former
appears to be more extended in R.A.\ and have slightly
different orientations than those in the latter. There is also
additional emission in 2007 to the south between the two lobes,
which is not observed in 2008. We believe that these discrepancies
are mostly due to artifacts in the image reconstruction, as
a result of sparse \uv\ sampling of the data \citep[see][]{hbk+09}. 
On the other hand, we argue that the two main lobes are physical, because
they are present in both observations which have different
hour angle coverage. The total cleaned flux density in the 2007 and
2008 images are 120 and 50\,mJy, respectively, much lower than
the values 350 and 410\,mJy obtained from ATCA monitoring
at a similar frequency in the same time period \citep{zsb+10}.
With a minimum baseline of $\sim$300\,k$\lambda$, the VLBI
observations are insensitive to features larger than $\sim0\farcs7$.
Hence, our result indicates that the majority of the remnant flux
density is contributed by smooth diffuse emission of at least a
comparable scale, implying a substantial thickness of the radio shell.

In Figure~\ref{img}b the 2008 VLBI image is compared to
lower resolution ATCA images at 9 and 36\,GHz taken at similar
epochs (2008 October) \citep{ngs+08,psn+09}. Since all these
observations are phase-referenced to the same nearby calibrator,
we estimate that the astrometric accuracy should be $\sim10$\,mas
and therefore did not perform any manual alignment. The lobes
found in the VLBI images closely match the brightest part of the
shell at 9 and 36\,GHz. However, regions of lower brightness
temperature tracing the remainder of the remnant, seen in the
latter two images, are not detected in the VLBI observations. This
is likely because the large-scale diffuse emission is resolved
out in the VLBI images, and any compact features (see
\S\ref{compact} below) fall below the detection sensitivity.
Figure~\ref{img}b also presents a comparison of the VLBI result
to H$\alpha$ and X-ray images obtained from the \emph{HST}
and the \emph{Chandra X-ray Observatory} on 2007 May and 2008
April, respectively \citep[J.C.S.\ Pun, private
communication;][]{ngm+09}. These images were fitted to a simple
torus to determine the remnant center, then aligned with the
VLBI image using the supernova position reported by \citet{psn+09}.
There appears to be no detailed correlation of the brightness
distribution at small scales in the radio, optical and X-ray bands.
Moreover, the images indicate that the radio shell has a
comparable size to the optical inner ring and the X-ray
shell. The latter result is consistent with the findings
by \citet{ngm+09}.

\subsection{Fourier Modeling}
\label{fourier}

\begin{figure*}[ht]
\epsscale{1.0}
\plotone{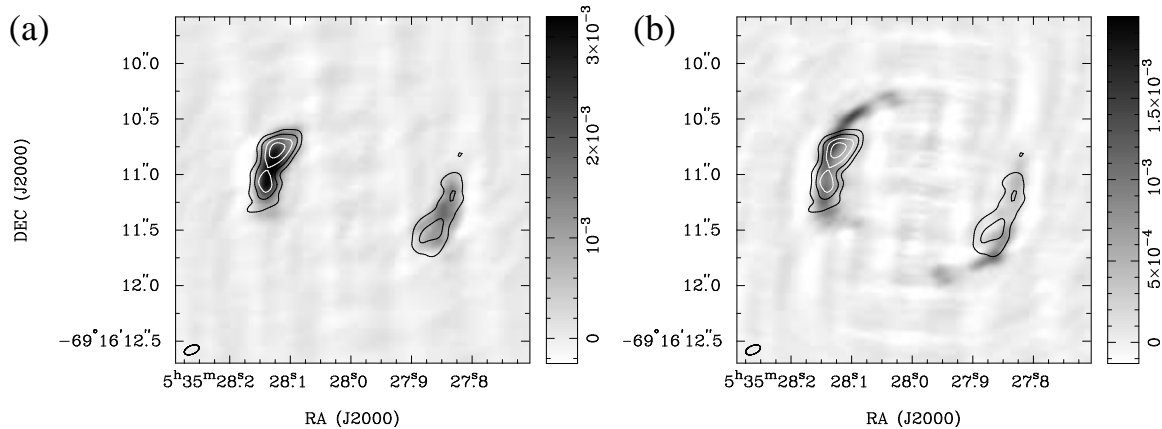}
\caption{Images made from simulations of the 2008 VLBI 
visibility data, based on the best-fit truncated-shell models
on the 2008 VLBI data (left) and on the 9\,GHz ATCA data (right).
The \uv\ sampling of the simulations are identical to the 
2008 VLBI observation. The images are noise-free, and the
contours are from the 2008 VLBI image in Figure~\ref{img}.
\label{sim}}
\end{figure*}

To quantify the remnant geometry, we employed the Fourier modeling
technique described by \citet{ngs+08} to fit the VLBI visibility
data with a truncated-shell model tilted at 43\fdg4 to the
line-of-sight \citep{pun07}. The model parameters include the center
position, flux, mean shell radius (averaged between the inner and
outer shells), half-opening angle, shell thickness, and a linear
gradient in the surface emissivity that accounts for the observed
asymmetry \citep[see Figure~3 in][for a detailed definition of all
these parameters,]{ngs+08}. Given the limited \uv\ coverage of the
data, we fixed the gradient using the 9\,GHz results \citep{psn+09}
in order to improve the fit stability. The level of gradient is fixed
at 36\% (ratio between the surface brightness at the edge and at
the center) with a position angle at 109\arcdeg\ (measured from the
observer's line-of-sight projected on the shell's equatorial plane).
All other parameters, including position, flux, radius, half-opening
angle and thickness, are free to vary in the fit. The $\chi^2$ values
are calculated by weighting the fit residuals with the measurement
uncertainties of each data point, which are estimated from the scatter
of the visibility amplitudes among the four channels, because
information on individual measurement errors were not available.
Table~\ref{table} summarizes the main results: the best-fit models
have radii 0\farcs82 and 0\farcs84 in 2007 and 2008, respectively,
half-opening angle $\sim20\arcdeg$ and thickness $\sim20\%$ of the
radius. We note that the measured thickness should be considered only
as a lower limit, since the data are insensitive to large-scale structure.
The small reduced $\chi^2$ values suggest that the measurement
uncertainties are likely overestimated. To determine the confidence
intervals of the best-fit parameters, we scaled the uncertainties
such that the reduced $\chi^2=1$, and found very small formal
statistical errors. This implies that the systematic errors
dominate, which could be introduced by calibration errors or by
the weighting scheme of the fits. While the former is difficult
to estimate, we attempted to quantify the latter with simple
least-square fits (i.e.\ without weighting). The difference in the
model parameters give the uncertainties reported in Table~\ref{table}.

\begin{figure*}[ht]
\epsscale{0.85}
\plottwo{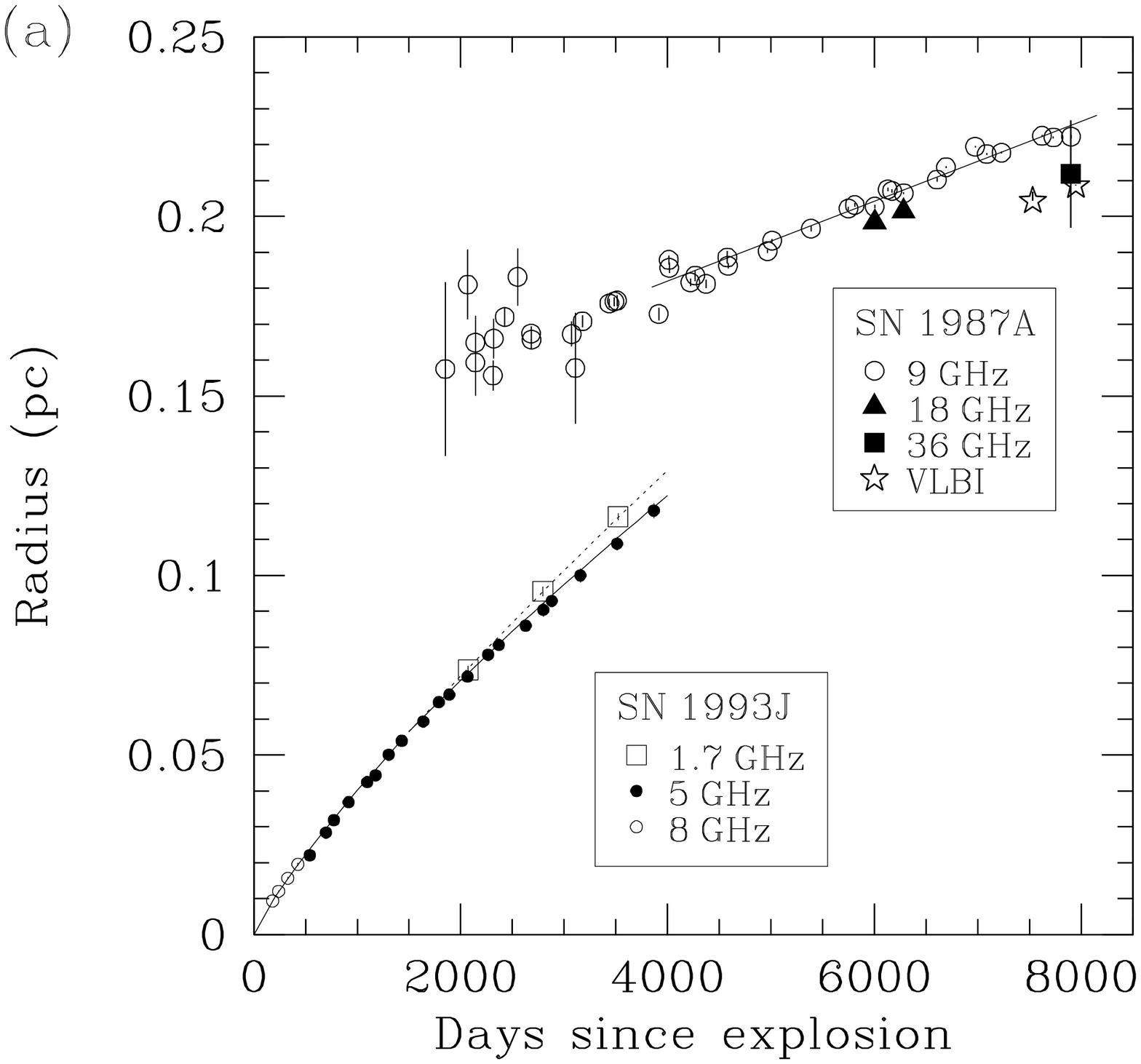}{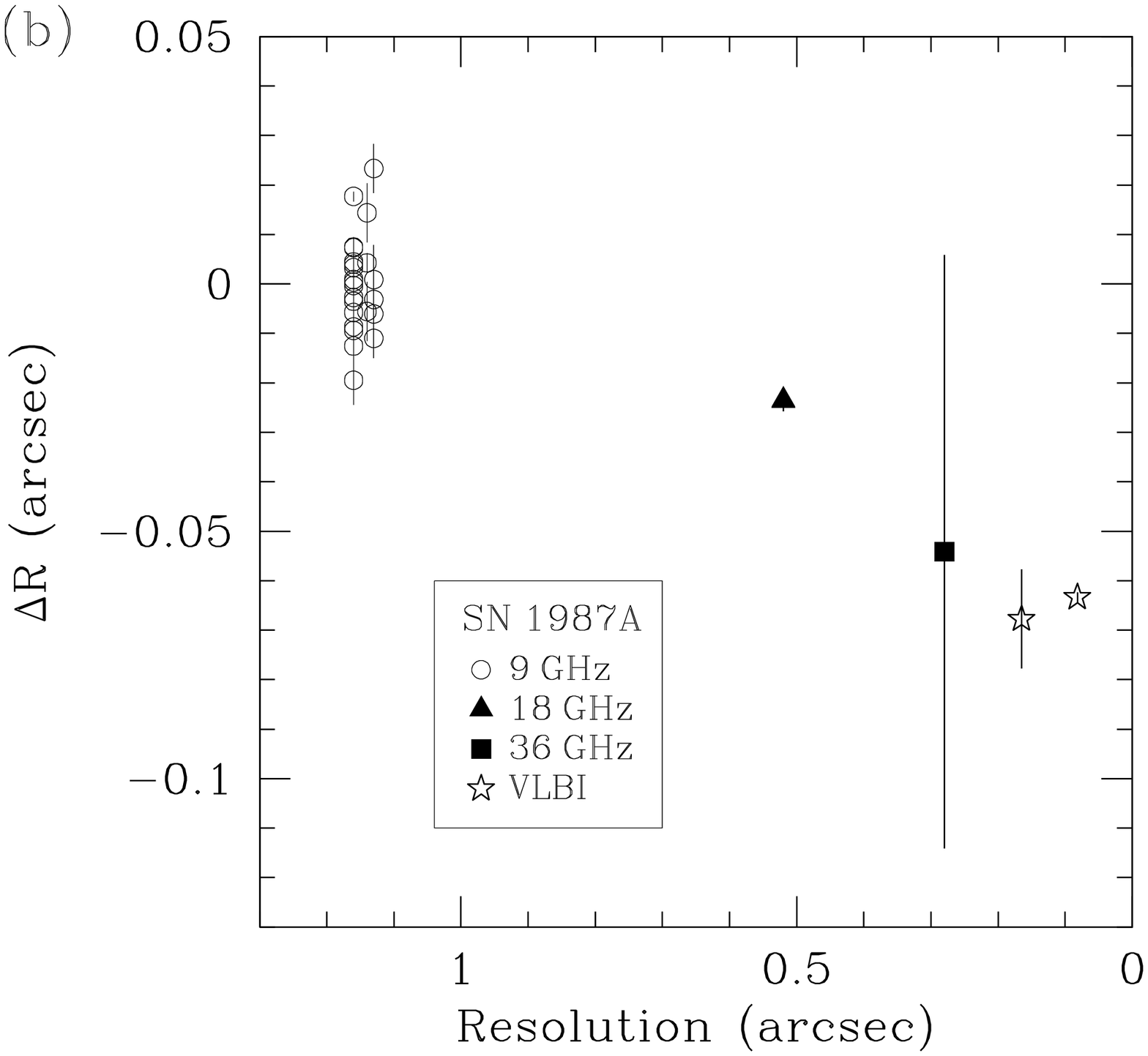}
\caption{
(a) Radii of supernova remnants 1987A and 1993J obtained from different
radio observations \citep{ngs+08,psn+09,mma+09}, assuming distances
of 51.4\,kpc and 3.96\,Mpc, respectively \citep{bbr+07}.
(b) Deviation of SN~1987A's shell radius versus the theoretical
resolution of different observations. The deviations are estimated
from a linear fit to the 9\,GHz radius beyond day 4000, which is
shown by the solid line in (a). \label{radius}}
\end{figure*}

Based on the best-fit model parameters for 2008, we simulated
a visibility data set using the \uv\ sampling of the 2008 VLBI
data. The simulated data were then imaged and deconvolved identically
as the real observations, and the resulting image is shown in
Figure~\ref{sim}a. The truncated-shell model successfully
captures the overall shell morphology, but fails to reproduce
the double-peak feature of the eastern lobe. A direct
comparison between the best-fit parameters in 2007 and 2008
indicates no significant change in the shell's geometry. The
difference in radius gives an expansion velocity of $4000\pm
3000$\kms, consistent with the result from \citet{ngs+08},
within the sizeable error due to the relatively short
time period between the observations. The VLBI images
show no obvious point source at the remnant center, and the 2008
VLBI image gives a 3$\sigma$ upper limit of 0.3\,mJy\ on the flux
density of any possible central object, such as a radio pulsar or
pulsar wind nebula at 1.7\,GHz.

As listed in Table~\ref{table}, the best-fit radii are
significantly smaller than the value 0\farcs89 obtained
from the ATCA 9\,GHz observations in 2008 \citep{psn+09}.
To determine if the discrepancy could be caused by the limited
\uv\ coverage of the VLBI data, we simulated an image with
the best-fit truncated-shell model of the 9\,GHz data, and
using the 2008 VLBI observations' \uv\ coverage. The result
is shown in Figure~\ref{sim}b, clearly indicating a larger
shell size than observed. This suggests that if the 9\,GHz
model were the correct description of the remnant geometry,
then the VLBI observations should be able to recover the
geometric parameters. We compare in Figure~\ref{radius}a
the mean shell radius obtained from Fourier modeling of
different radio observations \citep{ngs+08,psn+09}, and
found that the best-fit radius appears to be largest at 9\,GHz
and smallest at 1.4\,GHz, while the sizes at
18 and 36\,GHz are in between.

Discrepancy in shell radius at different frequencies has
been claimed for the radio remnant of SN~1993J, with a
progressively smaller radius at 5\,GHz than at 1.6\,GHz
since day 1500 \citep{mma+09}.\footnote{However, we note that
\citet{bbr+02} did not find such a correlation using the same
dataset.} A few possible explanations have been proposed,
including difference in electron synchrotron lifetimes, a radially
increasing magnetic field, or frequency-dependent changes in
the opacity of the medium \citep{mma+09}. For our case of
SN~1987A, the non-monotonic change of the mean shell radius with
frequency presents a more complex picture, suggesting that the
above physical explanations may not apply. In Figure~\ref{radius}b
we plot the deviation of radius from a linear fit to the
9\,GHz radius after day 4000 against the theoretical (i.e.\ 
diffraction-limited) angular resolution. We found a trend
of decreasing radius as the beam size gets smaller,
independent of the observing frequency. We have examined
whether biases in our fitting technique may give rise to
such a result, but as the fits are carried out in the
Fourier domain rather than the image domain, our procedure
is relatively immune to this. Therefore, we suggest that
measurements at higher spatial resolutions are biased by
unmodeled components in the radio emission, likely to be
bright knots near the shell's inner surface.

To test this idea, we tried applying the Fourier modeling to
the 18 and 36\,GHz ATCA data, but restricted on the \uv\ 
range $<180\mathrm{k}\lambda$, corresponding to the same
coverage as the 9\,GHz ATCA observations. In all cases,
this results in a $\sim5$\% increase in the best-fit radii
when compared to fits on the full dataset, giving values
closer to the 9\,GHz results, thus, providing further support
to the picture above. As a final check, we simulated visibility
data based on a toy model of a truncated shell plus few point
sources. The shell has a radius of 0\farcs9 and 10\% thickness,
and we added four point sources, two at each side, at a distance
0\farcs8 from the center and with a total flux density of 15\%
of the remnant. Based on this toy model, we simulated two visibility
data sets using the \uv\ sampling of the 2008 VLBI and ATCA
9\,GHz observations, then applied the Fourier modeling as in the
real observations. This gives radii of 0\farcs85 and 0\farcs88
for the two simulated data sets, respectively, confirming that
our radius measurements made with a truncated shell could be
biased by compact features in the remnant, especially at high
spatial resolution. This result is consistent with previous
findings \citep{mgw+02,ngs+08}, in which the radius is always
smaller when fitted with a single shell than with a shell plus
point sources. We note that although previous results
at 9\,GHz based on Fourier modeling \citep[e.g.][]{ngs+08} are
likely biased too, our simulations suggest that this is at a
much lower level than the VLBI case. Also, the bias is systematic,
therefore, it should have a minimal effect on the expansion
rate measurements.

\subsection{Small-scale Features}
\label{compact}
Results from the Fourier modeling above present evidence
of small-scale features in the radio remnant of SN~1987A. As
hinted by the double-peak structure of the radio lobes seen
in Figure~\ref{img}a, these features likely consist of discrete
clumps near the inner surface of the shell. To constrain their
flux density and spatial extent, we assume an extreme case that all
diffuse emission is resolved in the VLBI observations, such
that the structure seen in the images represents only the
compact emission. We fitted the 2008 VLBI visibility data with
a simple model of four Gaussian blobs, two at each lobe, and
found that the clumps have size smaller than $0\farcs2$,
with a distance 0\farcs78-0\farcs83 from the remnant center,
and a total flux density $<$\,50\,mJy. The size limit is only marginally
larger than the restoring beam, and converts to a physical
extent $<1.5\times10^{17}$\,cm ($\approx0.05$\,pc) at the distance
51.4\,kpc. This is comparable to the scale $<10^{17}$\,cm of
the optical hot spots, for which hydrodynamic simulations
suggest a physical size of $2\times10^{16}$\,cm, too small to
be resolved by the \emph{HST} \citep{pmz+02}. While
it is tempting to associate the radio clumps with the optical
hot spots, which are dense gas protruding inward from the
inner ring, Figure~\ref{img}b shows no correlation between
the radio and optical brightness distribution. Also, the first
optical hot spot emerged in the northeast with a position angle
29\arcdeg\ \citep[see inset of Figure~\ref{img}b;][]{lsb+00},
significantly offset from the radio peak. Alternatively, the
compact radio emission could indicate sites of enhanced
cosmic ray acceleration, either with magnetic field
amplification or with a local field orientation parallel to the
shock normal. Hence, tracing the evolution of the clumps
in future observations may give direct constraints on
the acceleration time scale and the magnetic field strength
\citep[see][]{uat+07}.

\section{Conclusion}
In this paper, we present a detailed study of VLBI observations
of SN~1987A taken in 2007 and 2008 at 1.4 and 1.7\,GHz, respectively.
This gives the first VLBI detection of the radio remnant, revealing
extended lobes that are consistent with the overall shell morphology
in previous radio studies. We found that the Fourier modeling
made with a simple truncated shell gives a smaller mean shell radius
for the VLBI data than for other radio observations at lower spatial
resolutions. This suggest small-scale features in the radio emission
that are not captured by the smooth model. These features are likely
to be discrete clumps near the shell's inner surface. We obtain a
limit of 0\farcs2 on their spatial scales, with a flux density limit
50\,mJy, corresponding to $\sim13\%$ of that of the remnant.

Although only part of the shell is detected in our VLBI
observations, this study demonstrated VLBI imaging as
a powerful tool to trace the fine structure of
the radio emission. Further observations utilizing the
full capability of the LBA, including more antennas and
wider bandwidth, will offer images with better sensitivity
and resolution. In addition, the on-going interaction
between the supernova blast wave and the dense CSM has
induced a rapid brightening of the remnant. Altogether,
we expect future VLBI images to reveal more clumps around
the shell, and potentially their substructure. The new
observations will also provide a direct measurement of the
remnant's expansion rate, tracing the shock evolution in the
CSM. Finally, we note that if in future the clumps start to
dominate the remnant flux, this will have a significant impact
on the Fourier modeling of the radio data. In particular, the
east-west asymmetry of the shell could depend sensitively
on the brightness of individual clumps, as a result, the simple
linear gradient approximation used by \citet{ngs+08} may
no longer hold.

\acknowledgements
We thank the referee for useful comments and we are grateful to
C.S.J.\ Pun and K.L.\ Li for providing the \emph{HST} images.
C.-Y.N.\ is a CRAQ postdoctoral fellow.
The Australian Long Baseline Array is part of the Australia
Telescope which is funded by the Commonwealth of Australia for
operation as a National Facility managed by CSIRO. e-VLBI
developments in Europe are supported by the EC DG-INFSO funded
Communication Network Developments project `EXPReS', contract
No.\ 02662 (\url{http://www.expres-eu.org}). The European VLBI
Network (\url{http://www.evlbi.org/}) is a joint facility
of European, Chinese, South African and other radio astronomy
institutes funded by their national research councils.

{\it Facilities:} \facility{LBA ()}, \facility{EVN ()}, \facility{ATCA ()}


\end{document}